\catcode`\@=11
\def\slash{\mathpalette\make@slash}
\def\make@slash#1#2{\setbox\z@\hbox{$#1#2$}%
  \hbox to 0pt{\hss$#1/$\hss\kern-\wd0}\box0}
\catcode`\@=12 
\newcommand{\eqnn}[1]{\begin{eqnarray*}#1\end{eqnarray*}\break}
\newcommand{\eqnl}[2]{\par\parbox{11cm}
{\begin{eqnarray*}#1\end{eqnarray*}}\hfill
\parbox{1cm}{\begin{eqnarray}\label{#2}\end{eqnarray}}\break}

\newcommand{\eqngrlb}[3]{\par\parbox{11cm}
{\begin{eqnarray}\fbox{$\displaystyle#1\\#2$}\end{eqnarray}}\hfill
\parbox{1cm}{\begin{eqnarray}\label{#3}\end{\eqnarray}}\break}

\newcommand{\eqngrl}[3]{\par\parbox{11cm}
{\begin{eqnarray*}#1\\#2\end{eqnarray*}}\hfill
\parbox{1cm}{\begin{eqnarray}\label{#3}\end{eqnarray}}\break}

\newcommand{\refs}[1]{(\ref{#1})}

\documentstyle[12pt]{article}

\advance\hoffset by -.35in
\begin{document}

\begin{flushright}DIAS-STP 95-30\\August 1995\end{flushright}

\def\pa{\partial}
\def\es{\!=\!}
\def\sg{\sqrt{g}}
\def\lam{\lambda}
\def\var{\varphi}
\def\La{\Lambda}

\def\A{{\cal A}}
\def\G{\Gamma}
\def\F{{\cal F}}
\def\J{{\cal J}}
\def\m{\mu}
\def\a{\alpha}
\def\b{\beta}
\def\p{\phi}
\def\g{\gamma}
\def\L{{\cal L}}
\def\H{{\cal H}}
\def\WW{{\bf W}}
\def\GG{{\bf \G}}
\def\hG{\hat\G}
\def\sqr#1#2{{\vcenter{\vbox{\hrule height.#2pt
        \hbox{\vrule width.#2pt height#1pt \kern#1pt
           \vrule width.#2pt}
        \hrule height.#2pt}}}}
\def\square{\mathchoice\sqr34\sqr34\sqr{2.1}3\sqr{1.5}3}

\begin{center}\Large{{Duality and the Legendre Transform}}
\end{center}
\vspace{1cm}

\begin{center}C.Ford and I.Sachs\\
{\it Dublin Institute for Advanced Studies,\\
10 Burlington Road, Dublin 4, Ireland.}
\end{center}

\vspace{2cm}

\begin{abstract}
We define a weak-strong coupling transformation based on the Legendre
transformation of the
effective action. In the case of $N\es 2$ supersymmetric
 Yang-Mills theory, this coincides
with the duality transform on the low energy effective action
considered by Seiberg and Witten. This Legendre transform interpretation
of duality generalizes directly to the full effective action, and in
principle to other theories.
\end{abstract}


In 1977 Montonen and Olive (MO) \cite{MO} proposed that the strong
coupling regime of
certain Yang-Mills-Higgs theories with gauge group $G$,
spontaneously broken to some compact subgroup $H$ is equally well
described by a
dual weakly coupled theory, where topological and Noether charges exchange
 roles. MO duality is believed to operate in $N\es 4$ Yang-Mills
(YM) theory
 \cite{Osb,WV}.
 Recently Seiberg and Witten have constructed a version of MO duality
 on the light fields of $N\es 2$ YM theories \cite{SW1,SW2}.

In this letter we observe that for $N\es 2$ theories the low energy effective
action of the dual theory is the Legendre transform of the ordinary low energy
effective action. That is, the dual effective action is the Schwinger
functional for some (topological) current.
This definition of the $S$ duality transformation
directly
extends to the
full effective
action, and in principle to other theories. Here we address some issues
associated with the
full effective potential in $N\es 2$ YM. We also discuss why this relation does
not hold in
the Sine-Gordon Thirring duality \cite{Col}.\par

To see how the dual low energy effective action in pure $N\es 2$ YM
with gauge group $SU(2)$ \cite{SW2}
is related to the generating functional for the topological current, couple a
source term
 \eqnl{\frac{1}{4\pi}
\hbox{Im}\int d^4x
d^2\theta_1 d^2\theta_2\; \A^a\J^a}{sf1}
to the classical action. Here $\A^a,\;a\es 1,2,3$ is a $N\es 2$ chiral
superfield, which satisfies a Bianchi
constraint  (see for example \cite{West}),
and
 $\J^a$ is a  $N\es 2$ chiral superfield. In component field
language the $\A^a\J^a$ term contains a coupling $ V_\mu j^\mu_{top}\es
V_\mu\epsilon^{\mu\nu\rho\lambda}
\pa_{\nu}F_{\rho\lambda}$ to the abelian topological current.
The Schwinger functional $\WW[\J^a]$ obtained from \refs{sf1} then leads to the
effective action $\GG[\A^a]$ after Legendre transformation\footnote{In order to
define the effective action, $\GG[\A^a]$,
 one must impose
a constraint on $\J^a$ (which matches the Bianchi constraint
on $\A^a$) chosen so that $\delta \WW[\J^b]/\delta
\J^a=\A^a$ is invertible.}.
In ref. \cite{SW1,SW2} attention was focussed on the \sl low energy
\rm effective action $\G[\A]$, where $\A$ refers to the light fields
($\A=\A^3$, say) and derivative terms (see e.g. \cite{Hen}) are ignored.
The low energy  effective action for pure
        $N=2$ $SU(2)$ YM
can be written in $N=2$ superspace \cite{Sei}
as follows
\eqnl{\G[\A]=\frac{1}{4\pi}\hbox{Im}\int d^4x
d^2\theta_1 d^2\theta_2 \F(\A),}{eff1}
where  $\F$ \footnote{ We can regard $\F$ as  a N=2 superspace effective
potential, although it will contain space-time derivative terms when
expanded out in component fields.} is a holomorphic
function. In ref. \cite{SW1} an exact (although implicit) expression was
given for $\F(\A)$, and it was argued that the theory has an equivalent
description in terms of the \lq\lq dual'' variables
\eqnl{\A_D=8\pi i{\delta\G\over{\delta\A}}={\partial\F(\A)\over{
\partial\A}},\quad
\bar \A_D=-8\pi i {\delta\G\over{\delta\bar\A}}={\partial\bar\F(\bar\A)
\over{
\partial\bar\A}},}{sd1}
in which the the magnetically charged solitons are treated as local fields. The
dual (low energy) effective action is now $4\pi\Gamma_D=\hbox{Im}\int
\F_D(\A_D)$ where the dual  potential $\F_D$ satisfies (see \cite{Hen} for
a $N\es 2$ formulation)
\eqnl{{\partial^2\F(\A)\over{\partial \A^2}}=
-\left({\partial^2\F_D(\A_D)\over{\partial\A_D^2}}\right)^{-1},}{sd2}
and $\hbox{Im}\F(z)$ is a convex function.
Using \refs{sd1} and \refs{sd2} it is  easy to see that an
 equivalent definition of
 $\G_D(\A_D)$ is given by
\eqnl{\G_D(\A_D,\bar\A_D)={}^{\hbox{min}}_{\,\A,\bar\A}\left[
 \Gamma[\A]-{1\over{4\pi}}\hbox{Im}\int \A_D\A\right].}{sd3}
Having established that $\G_D(\A_D)$ is the
Legendre transform of $\G(\A)$, it follows at once that it must be
(the convex hull of) the low energy
generating functional obtained from \refs{sf1}.
Of course the Schwinger functional is a well defined object for any
theory, and in particular the so defined duality
transformation extends immediately to the full effective action for $N\es 2$
YM. We now examine the full superfield effective potential.
A gauge invariant extension of the low energy effective potential proposed
in \cite{SW2,Sei} is given by
\eqnl{ \H(\A^a\A^a)=\F(\sqrt{\A^a\A^a}),}{sh1}
and so one can write down a gauge invariant extension of the low energy
effective
action as\footnote{This is still not the full effective action, since we ignore
higher derivative terms.}
$4\pi\hG\es\hbox{Im}\int\H$.
As pointed out in \cite{Roc}
\eqnl{\hbox{Im}{\partial^2\H(\A\cdot\A)\over{\partial\A^i\partial\A^j}}
     =\hbox{Im}\Biggl[\F''(\sqrt{\A\cdot\A}){\A^i\A^j\over{\A\cdot\A}}+
       {\A_D\cdot\A\over{\A\cdot\A}}\left(
       \delta^{ij}-{\A^i\A^j\over{\A\cdot\A}}\right)\Biggr],}{sh2}
is not positive definite in a certain region of the moduli space.
On the other hand, we know that the effective action should be convex. So
it seems the high energy effective potential given by \refs{sh1} does not
properly describe this region. One possibility is to replace $\hG$ by its
convex hull. We now define the dual high energy effective potential as
in \refs{sd3} by
\eqnl{\hG_D(\A_D^a,\bar\A_D^a)={}^{\hbox{min}}_{\,\A^a,\bar\A^a}\left[
 \hG[\A^a,\bar\A^a]-{1\over{4\pi}}\hbox{Im}\int \A_D^a\A^a\right],}{sh2.1}
where it is understood that $\A^a$ satisfies the Bianchi constraint.
Note that $\G_D$ is automatically convex, since we have defined it
as a Legendre transform.
In
 the region(s) of the moduli space where the Hessian \refs{sh2}
 is positive definite, a solution to \refs{sh2.1} is given by
\eqnl{\hG_D(\A_D^a,\bar\A_D^a)={1\over{4\pi}}\hbox{Im}\int
d^2\theta_1 d^2\theta_2 \H_D(\A^a_D\A^a_D),}{sh2.2}
where
\eqnl{\H_D(\A^a_D\A^a_D)=\F_D(\sqrt{\A^a_D\A^a_D})}{bill}
and
\eqnl{\A_D^a={\partial \H(\A\cdot\A)\over{\partial\A^a}}=
{\F'(\sqrt{\A\cdot\A})\A^a\over{\sqrt{\A\cdot\A}}}.}{ben}
In the  region where the matrix \refs{sh2} is not positive
definite we do not have an explicit expression for the dual effective action,
but its
stability is guaranteed by general properties of the Legendre transform.\par
Seiberg and Witten have also extended their work to $N\es 2$ gauge
theory with $N\es 2$ matter multiplets \cite{SW2}. In this case the matter
fields seem to play a passive role in the duality. Therefore in
order to obtain the dual effective action one would only Legendre
transform with respect to the $N\es 2$ gauge fields.
\par

The situation in the Thirring Sine-Gordon duality \cite{Col} is
somewhat different.
 This duality leads to an identity between the  Schwinger
functionals for the respective conserved currents.
 Indeed, proceeding as in \cite{Col} one can show that
\eqngrl{\exp\left(i\WW_T[J_\m]\right)&=&
 \int d\bar\psi d\psi \exp\left(
 i\int d^2x\left[\L_T+J_\m\bar\psi\g^\m\psi\right]\right)}
 {&=&\int d\p\exp\left(
 i\int d^2x\left[\L_{SG}+J_\mu j^\mu_{top}\right]\right) =
 \exp\left(i\WW_{SG}[J_\mu]\right),}{th1}
where\footnote{Our notation follows that of \cite{Col}} $2\pi j^\mu_{top}\es
-\beta\epsilon^{\mu\nu}\pa_\nu\p$ and identification of the
couplings is made as in \cite{Col}. Hence in contrast to $N\es 2$ YM, the
respective Schwinger functionals are identical. If there was to be a
Legendre transform interpretation of Coleman duality we would have the \lq\lq
self duality'' relation $\WW_{SG}[J_\mu\es
2\pi\epsilon_{\mu\nu}\partial^\nu\Phi/\beta]=
\GG_{SG}[\Phi]$, where $\GG_{SG}[\Phi]$ is the usual effective action for the
Sine-Gordon model. This is certainly not expected. Up to second order in
perturbation
 theory we observed
\eqnn{\WW_{SG}[J_\mu=
2\pi\epsilon_{\mu\nu}\partial^\nu\Phi/\beta]=\GG_{SG}[\Phi]-\int d^2x
\partial_\m\Phi\partial^\m\Phi.}
The Thirring Sine-Gordon duality relates the Schwinger functional for
 the Noether current of the Thirring
model with the same for the topological current in the Sine-Gordon model,
whereas the $N\es 2$ YM duality relates the Noether and topological
sectors of the same theory.
\par

To summarize, we have shown that in $N\es 2$ YM theory
the low energy effective action for the dual theory is the Schwinger
functional for the topological current of the original theory.
 Here we have verified this weak-strong
 coupling relations for the low-energy effective
 action of $N\es 2$ YM, although it might
well extend directly to the full
effective action as well as other theories. In particular, it would be
interesting to see how these ideas extend to N=1 theories. If such
a relation exists, it would reduce the computation of the S-matrix
 in the strongly coupled theory to the weak
coupling expansion of the dual effective action.\par
\par
\vspace{1cm}
 We are grateful to G.A.F.T. da Costa, B.Dolan, D.O'Connor and
 L.O'Raifeartaigh for numerous discussions on ref. \cite{SW1} and
C. Wiesendanger for a careful reading of the manuscript.

\end{document}